# Searching for Protoplanets around MWC 758 and MWC 480 in Br-γ Using Kernel Phase and SCExAO/CHARIS

Alexander Chaushev[1], Steph Sallum[1], Julien Lozi[2], Jeffrey Chilcote[3], Tyler Groff[4], Olivier Guyon[2,5,6],
N. Jeremy Kasdin[7], Barnaby Norris[8], and Andy Skemer[9]
[1] Department of Physics and Astronomy, University of California Irvine, 4129 Frederick Reines Hall, Irvine, CA 92697, USA; a.chaushev@uci.edu
[2] Subaru Telescope, National Astronomical Observatory of Japan, 650 North A'ohōkū Place, Hilo, HI 96720, USA
[3] Department of Physics, University of Notre Dame, 225 Nieuwland Science Hall, Notre Dame, IN 46556, USA
[4] NASA-Goddard Space Flight Center, Greenbelt, MD, USA
[5] Astrobiology Center, National Institutes of Natural Sciences, 2-21-1 Osawa, Mitaka, Tokyo, Japan
[6] Steward Observatory, University of Arizona, 933 N Cherry Avenue, Tucson AZ 8572, USA
[7] University of San Francisco, San Francisco, CA 94118, USA
[8] Sydney Institute for Astronomy, School of Physics, University of Sydney, Sydney, NSW, Australia
[9] Department of Astronomy & Astrophysics, University of California, Santa Cruz, CA 95064, USA
Received 2023 October 5; revised 2024 May 21; accepted 2024 May 29; published 2024 July 11

## Abstract

Discovering new actively accreting protoplanets is crucial to answering open questions about planet formation. However, identifying such planets at orbital distances where they are expected to be abundant is extremely challenging, both due to the technical requirements and large distances to star-forming regions. Here we use the kernel phase interferometry (KPI) technique to search for companions around the ∼6 and ∼8 Myr old Herbig Ae stars MWC 758 and MWC 480. KPI is a data analysis technique that is sensitive to moderate asymmetries, arising from, e.g., a circumstellar disk or companions with contrasts of up to 6–8 mag, at separations down to and even below the classical Rayleigh diffraction limit ($\sim 1.2\lambda/D$). Using the high-spectral-resolution $K$-band mode of the SCExAO/CHARIS integral field spectrograph, we search for both excess Br-γ line emission and continuum emission from companions around MWC 480 and MWC 758. We are able to set limits on the presence of rapidly accreting protoplanets and brown dwarfs between 4 and 16 au, well interior to those of previous studies. In Br-γ, we set limits on excess line emission equivalent to accretion rates ranging from $10^{-5} M_j^2 \cdot \mathrm{yr}^{-1}$ to $10^{-6} M_j^2 \cdot \mathrm{yr}^{-1}$. Our achievable contrasts demonstrate that KPI using SCExAO/CHARIS is a promising technique to search for giant accreting protoplanets at smaller separations compared to conventional imaging.

*Unified Astronomy Thesaurus concepts:* Exoplanet formation (492); Protoplanetary disks (1300); Young stellar objects (1834); Interferometry (808); Astronomy data analysis (1858); High contrast techniques (2369); Spectroscopy (1558)

## 1. Introduction

Despite the abundance of discovered exoplanets, many open questions remain about the formation processes that have produced them. Some fundamental questions include how quickly planets grow, how the rate of growth changes over time, and where the most common formation sites are found within the circumstellar disk. A major hurdle in addressing these questions is the difficulty of directly studying young accreting protoplanets. While the contrast ratios needed to detect protoplanets are relatively modest, on the order of $10^{-4}$ (Eisner 2015; Zhu 2015), this is challenging to achieve at the tight angular separations required by the large distances to nearby star-forming regions (Gaia Collaboration et al. 2018; Luhman 2023). For example $\lambda/D$ in $K$ band at 150 pc corresponds to ∼10 au for current 8 m class telescopes, twice the orbital separation of Jupiter. Finally, identifying protoplanets around a young star presents its own challenges; for example, extinction from the ample dust in the environment may limit sensitivity, and the complex structure of the circumstellar disk itself may make it difficult to unambiguously identify a planet (e.g., LkCa 15; Sallum et al. 2023).

So far, only a handful of young candidate protoplanets have been identified, with only PDS 70 b and c being unambiguously confirmed (Keppler et al. 2018; Müller et al. 2018; Haffert et al. 2019). A study of the PDS 70 system by Haffert et al. (2019) has reported Hα line emission for both protoplanets. Haffert et al. (2019) used an Hα line emission search to identify PDS 70 c and confirm the planetary nature of PDS 70 b. This demonstrates that accretion signatures, including Br-γ, can be a powerful tool for identifying planetary companions. Furthermore, the PDS 70 b and c Hα fluxes indicated low accretion rates, which would form Jupiter-mass planets in 50–100 Myr (much greater than the expected disk lifetime). Given that the age of PDS 70 is ∼5 Myr (Müller et al. 2018), this suggests that PDS 70 b and c represent the end stages of the formation process where the majority of accretion has already taken place. However, Hα is not the only accretion tracer available. Recent near-infrared detections of Paβ, Paγ, and Br-γ in the spectrum of the circumbinary planetary mass companion Delorme 1 (AB)b point to the possibility of using other lines to constrain accretion properties (Betti et al. 2022). The same study also detects evidence of short-term fluctuations in the accretion-line luminosity, which may imply variability in accretion physics such as infall rate and geometry and/or quickly varying extinction.

These studies have given us the first direct glimpses of planet formation. However, in order to fully understand the process, it is necessary to build up a larger census of actively accreting planets. One way of achieving this would be to search for protoplanets on orbits in the range of 5–10 au closer to the host star, where planet formation is expected to be more efficient. For example Delorme 1 (AB)b has a separation of 84 au, while the PDS 70 b and 70 c planets are at approximately 20 and 35 au, respectively. Evidence from high-angular-resolution imaging studies, such as the GPIES survey (Nielsen et al. 2019), has shown that the occurrence rate of Jovian mass planets and substellar companions is expected to peak below 10 au. Accessing regions below 10 au is exceptionally challenging for most coronagraphic instruments, which are typically limited by quasistatic speckles at 1–3 $\lambda/D$ (Ruane et al. 2019). New imaging methods are needed to allow access to this parameter space.

Finally, long(er)-baseline interferometry using the Large Binocular Telescope Interferometer (LBTI), Very Large Telescope Interferometer (VLTI), or Center for High Angular Resolution Astronomy (CHARA) array can provide improved angular resolution over a conventional single-dish telescope. This is made possible by leveraging the short spatial frequencies made accessible by longer baselines in the arrays. For example, the LBTI can currently reach Extremely Large Telescope-level angular resolution by using the fixed 23 m baseline between the two mirrors, while the VLTI and CHARA arrays can reach down to milliarcsecond separations using greater-than-100 m baselines. This allows for direct detection of exoplanets as well as indirect searches from astrometry. However, in practice, long-baseline interferometry is extremely challenging due to the difficulty of fringe tracking on fainter stars. In the future, this limit may be improved, reaching as low as $K = 22$ in the case of the future VLIT/GRAVITY+ instrument (Gravity+ Collaboration 2022). For long-baseline interferometry, in general, improving sensitivity is a matter of boosting adaptive optics (AO) performance using better hardware or improved understanding of the effects of the atmosphere (e.g., P-REX; Widmann et al. 2018; Perera et al. 2022).

We recently demonstrated the use of the kernel-phase interferometry technique (KPI; Martinache 2010; Martinache et al. 2020) with the Coronagraphic High Angular Resolution Imaging Spectrograph (CHARIS) instrument on the Subaru Telescope (Chaushev et al. 2023). CHARIS is an integral field spectrograph (IFS) that sits behind the SCExAO extreme AO system (Groff et al. 2015, 2016; Brandt et al. 2017). KPI is a data analysis method sensitive to close-in asymmetries possibly arising from structures such as disks or planetary companions. KPI allows for the achievement of similar angular resolutions as aperture masking (Tuthill et al. 2000) but without the corresponding loss of throughput (Sallum & Skemer 2019). By using CHARIS in the high-resolution $K$-band mode, we can search for rapidly accreting protoplanets that are bright in Br-$\gamma$, and additionally we can search for continuum emission in the $K$ band from more massive brown dwarf companions. In this way, we can jointly detect and characterize any potential planet candidates.

We use CHARIS KPI to study two young stars, MWC 480 and MWC 758. MWC 480 is a $6.2^{+0.31}_{-1.1}$ Myr Herbig Ae star found at $161.8^{+3.4}_{-3.2}$ pc (Vioque et al. 2018). The circumstellar disk around MWC 480 is host to a wealth of structures including two tight spirals at 162 au (1″) and 245 au (1″.5), two gaps at 73 au (0″.46) and 141 au (0″.87), as well as other spirals, temperature minima and maxima, and rings (Teague et al. 2021). Gaps, spirals, and other asymmetries can be powerful dynamical tracers of a planet, providing indirect evidence of its properties. Liu et al. (2019) performed global, smoothed 3D hydrodynamic simulations of an accreting planet in a circumstellar disk and found that the dust emission could be explained by an embedded $\sim 2.3\,M_{\rm jup}$ planet at $\sim 78$ au. Later simulations performed by Teague et al. (2021) suggest that many of the features observed in the disk could be driven by a $\sim 1\,M_{\rm jup}$ planet located farther out at $\sim 245$ au. As such, the rich structure of the disk makes MWC 480 a prime hunting ground for young accreting planets.

The other target is MWC 758, also a Herbig Ae star, that is $8.3^{+0.41}_{-1.4}$ Myr old at $160.3^{+2.9}_{-2.8}$ (Vioque et al. 2018). In addition to hosting a complex circumstellar disk, it also has two reported candidates in $L$ band, one at 20 au inside the submillimeter cavity (Reggiani et al. 2018) and another at 100 au (Wagner et al. 2019), with the latter being confirmed using the LBTI (Wagner et al. 2023). Both objects are likely planetary with a mass range of 5–8 $M_{\rm jup}$ though there is considerable uncertainty in this determination due to the unknown levels of extinction. Using an age estimate of 1.5–5.5 Myr, Boccaletti et al. (2021) were able to set upper-mass limits on both candidates of $\sim 5$ and $\sim 8\,M_{\rm jup}$, as well as search for other candidates down to $\sim 0″.15$. Boccaletti et al. (2021) performed a search using Very Large Telescope (VLT) SPHERE instrument to try and identify counterparts in $H$ and $K$ bands but failed to find any, even after adjusting for potential differences in extinction.

With KPI, we are able to extend this search down to well below 0″.05 in $K$ band (albeit at more moderate contrasts).

This study presents a high-angular search for companions around the systems MWC 480 and MWC 758, using KPI and SCExAO/CHARIS. The paper is divided as follows: in Section 2 we present the observations and archival data; in Section 3 we present the data analysis method for producing kernel phases; in Section 4 we present the results including calibration, Br-$\gamma$ and continuum emission search; and finally, in Section 5 we present the conclusions and future work.

## 2. Observations

We analyze two data sets for this study: new observations of MWC 758 taken in March of 2021 and previously unpublished archival data of MWC 480 taken in February of 2019. Both data sets consist of SCExAO/CHARIS high-resolution $K$-band observations. This mode has an average spectral resolution of $R = 77.1$ and 17 output channels with central wavelengths ranging from 2015 to 2368 nm.[10] The images have a 2″.07 by 2″.07 field of view and a spatial scale of 16.2 mas per lenslet. CHARIS sits behind the SCExAO/AO188 extreme AO system, which provides the high-Strehl AO correction necessary for KPI.

For MWC 480, the data were taken from the SMOKA Science archive.[11] The observations were conducted on the 2019 February 25. The data set consists of 709 frames for MWC 480 and 55 belonging to a point-spread function (PSF) reference star HD 56386. The reference star was dithered

---

[10] CHARIS covers all of the $J$, $H$, or $K$ band in a low-resolution mode or a single band at higher resolution.
[11] https://smoka.nao.ac.jp/

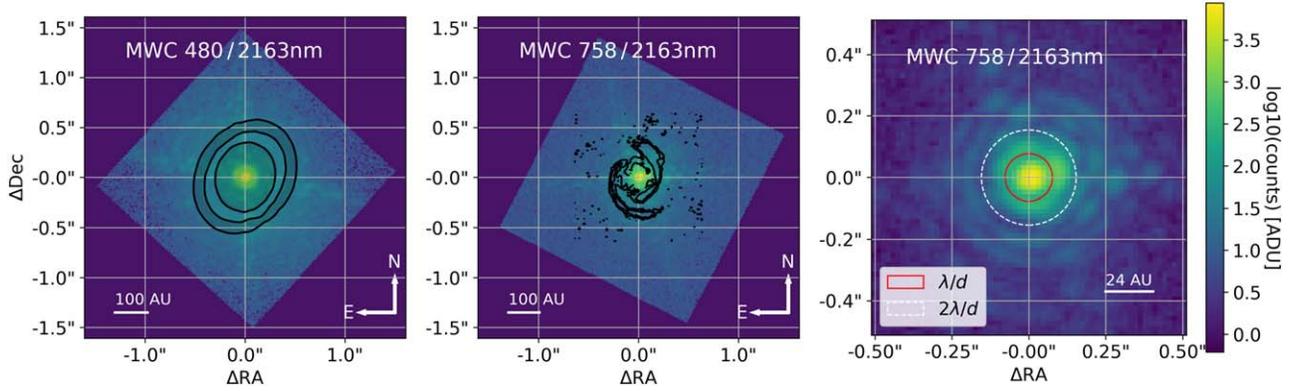

**Figure 1.** Full-frame example images from the MWC 480 (left panel) and MWC 758 (middle panel) CHARIS high-resolution *K*-band data sets, at the Br-γ spectral bin. The black contours and shaded region on the MWC 458 image are taken from 1.3 mm ALMA continuum observations (Liu et al. 2019) and show the distinct ring structure around the inner disk of MWC 480. The black lines on the MWC 758 image show contours from a Stokes $Q_\phi$ image taken with VLT/SPHERE (Ren et al. 2020). These contours show the distinct spiral arms of the system. The panel on the right shows a 64-by-64 pixel cutout of the MWC 758 image, the same size as used in the kernel-phase analysis. Two lines are shown for scale, one at a radius of $\lambda/D$ (red solid) and the other at $2\lambda/D$ (white dashed). Using a maximum baseline length of 5.8 m (see Section 3 for a discussion), a λ of 2163 nm gives a $\lambda/D$ of 77 mas, corresponding to 12 au at 160 pc or just under 5 pixels on the detector.

**Table 1**
Summary of the MWC 758 and MWC 480 Data Sets Along with PSF Reference Calibrators

| Identifier | Type | Obs-date | Gaia $R_p$ | *H* Band | *K* Band | $t_{frame}$ (s) | $n_{frames}$ | $n_{used}$ (%) | ΔPA (deg) |
|---|---|---|---|---|---|---|---|---|---|
| MWC 758 | Science Target | 2021-03-19 | 8.1 | 6.6 | 5.8 | 20 | 129 | 93.7 | 23.0 |
| HD 245009 | PSF Calibrator | 2021-03-19 | 7.5 | 6.0 | 5.8 | 20 | 42 | 95.2 | 1.2 |
| MWC 480 | Science Target | 2019-02-25 | 7.8 | 6.3 | 5.5 | 13 | 709 | 62.4 | 36.0 |
| HD 56386 | PSF Calibrator | 2019-02-25 | 6.2 | 6.2 | 6.2 | 10; 6 | 55 | 27.3 | 0.4 |

**Note.** The number of frames reflect the size of the full data set; however, only a subset of those frames are used for the subsequent analysis. The ΔPA is the difference in parallactic angle between the first and last observations. The final ΔPA of the data used for the analysis is smaller due to gaps in coverage (MWC 758) and the removal of many high-noise frames (MWC 480). The first three frames of the HD 56386 data set have a higher exposure time of 10 s versus 6 s for the remainder of the observations. The kernel phases from these frames were manually inspected to confirm that the longer exposure time was not an issue.

across several positions. Flat-field and dark frames were also available in the archive. Seeing information is not available for the night; however, the estimated Strehl ratio ranges from 0.3 to 0.8, indicating either variable seeing or AO performance. The Strehl was estimated by comparing the flux ratio inside and outside of a 0.″5 aperture to that of a model PSF computed with the Poppy library (Perrin et al. 2012). This gives a relative indication of the data quality throughout the night; however, the value may be subject to systematic biases arising from an imperfect estimation of the model PSF. Figure 1 shows two typical images for the Br-γ spectral bin from each data set. Table 1 contains the details of both data sets, including the two science targets and their respective calibrators.

The MWC 758 observations were conducted on the 2021 March 19 using a quarter night of time. MWC 758 is a bright target, which allows for a high-quality AO correction necessary to use the kernel-phase technique (Strehl > ∼ 0.8). To avoid saturating the target, a 10:90 beam splitter was used with 10% of the light sent to CHARIS and the remaining going to the wave-front sensor. The data set consists of 129 frames for MWC 758 and 42 frames for a PSF calibrator HD 245009. HD 245009 was chosen with the SearchCal tool (Bonneau et al. 2006, 2011). SearchCal estimates the stellar diameter of a star, given a spectral type and VJHK magnitudes, by modeling of the relationship between known angular diameters and photometries (Chelli et al. 2016). Additionally, both HD 245009 and HD 56386 were vetted using KPI, with all evidence pointing to them being single stars. Dark and flat-field calibrations were taken at the start of the night. No seeing information is available for the observations; however, the Strehl ratio was estimated to be in the range 0.75–0.85 throughout the night.

### 3. Kernel-phase Data Processing

KPI is a data processing technique where a conventional telescope is treated as an interferometric array (Martinache 2010; Martinache et al. 2020). In doing so, KPI allows for the detection of asymmetries in the source brightness distribution close to and even inside $\lambda/D$, which can correspond to a companion or features in a circumstellar disk. KPI has been used successfully for a number of science cases including studying brown dwarfs (Pope et al. 2013; Factor & Kraus 2023) and the prototypical pre-main-sequence system T Tauri (Kammerer et al. 2021). Since KPI is a data processing method, it can also be used to extend the capability of existing instruments such as JWST/ NIRISS (Kammerer et al. 2023). Here we summarize the theory behind kernel phases and the data processing pipeline used to extract them. The data processing pipeline used to derive kernel phases from the CHARIS data cubes is detailed in Chaushev et al. (2023).

The pupil is modeled as a dense interferometric array and discretized into a set of virtual subapertures, each of which can have a transmission varying from 0 to 1. The pupil plane phases and the phases measured from the Fourier-transformed image can then be linked via the following equation:

$$\Phi(u, v) = \text{Arg} \sum_{ij} \exp^{i(\phi_j - \phi_i)} + \Phi_0, \quad (1)$$

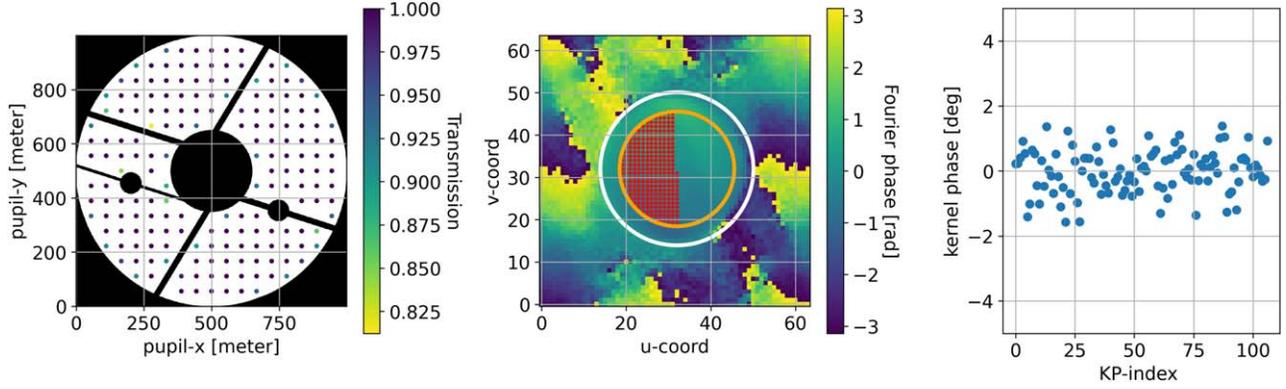

**Figure 2.** Steps in computing the kernel phases. Left panel: the discretized "gray" pupil model overlaid on the full SCExAO/CHARIS pupil. Middle panel: the UV plane of a single frame of HD 245009 computed with a discrete Fourier transform showing the phases and the sampling points used. The orange circle represents the 5.8 m baseline cutoff, while the white circle shows the full 7.74 m pupil. Right panel: example kernel phases from one frame of one of the PSF calibrators.

where $\Phi$ is the measured phase in the Fourier transform of the image, $\phi$ is the pupil plane phase, $\Phi_0$ is the intrinsic phase of the science target, and the indices $i$ and $j$ correspond to subapertures in the pupil model with spatial frequencies $(u, v)$. For small residual-phase errors, corresponding to a high-Strehl regime (e.g., where the AO correction is working well), a linear relationship between the pupil plane and Fourier phases can be derived by Taylor expanding the exponential term of each spatial frequency.

The measured Fourier phases can then be rewritten as

$$\Phi = R^{-1} \cdot A \cdot \phi + \Phi_0, \quad (2)$$

where $R^{-1} \cdot A$ describes how the pupil plane phase $\phi$ maps to the Fourier domain. The matrix $A$ relates the baselines to the subapertures, while the matrix $R^{-1}$ encodes the redundancy of each baseline.

Using singular value decomposition, we can then find a matrix $K$ such that $K \cdot A = 0$. Left multiplying Equation (2) through by $K \cdot R$ gives

$$K \cdot R \cdot \Phi = K \cdot R \cdot R^{-1} \cdot A \cdot \phi + K \cdot R \cdot \Phi_0 \quad (3)$$
$$= K \cdot A \cdot \phi + K \cdot R \cdot \Phi_0 \quad (4)$$
$$= K \cdot R \cdot \Phi_0, \quad (5)$$
$$= K_0 \cdot \Phi_0, \quad (6)$$

where $K_0$ is the kernel-phase matrix and $K_0 \cdot \Phi_0$ are the "kernel phases" with first-order perturbations eliminated. These perturbations can come from a number of instrumental sources, for example, from residual AO errors.

For the CHARIS data, the pupil model is discretized in subapertures with size 0.43 m. This give a total of 18 subapertures to fill the full diameter of the pupil. Each subaperture is given a transmission value as specified in Martinache et al. (2020), and the kernel-phase matrix is then computed using the XARA pipeline (Martinache 2010; Martinache et al. 2020).[12] Spatial frequencies corresponding to baselines above 5.8 m were removed prior to calculating the matrices in order to maximize the signal-to-noise ratio (SNR; Chaushev et al. 2023). As can be seen in Figure 2, the final pupil geometry results in 182 subapertures and 288 UV-sampling points with baselines less than the 5.8 m cutoff when using a 0.43 m grid spacing. Somewhat counterintuitively, removing baselines longer than 5.8 m greatly improves the SNR of the final kernel phases as was found in Chaushev et al. (2023). This is unexpected since longer baselines, which correspond to higher spatial frequencies, should improve the sensitivity of the kernel phases at small separations. However, in practice, these longer baselines are less coherent than the shorter baselines that have higher redundancy (within the context of the simplified KPI pupil model). The longer baselines, therefore, have lower SNR. Removing them and making use of the highly redundant smaller baselines instead boosts sensitivity.

To calculate the kernel phases, the CHARIS data were reduced using dedicated flat-field and dark current calibration frames that were taken at the start of observations. The data cubes were then extracted using the `CHARIS-DEP' pipeline (Brandt et al. 2017). After the cubes have been created, the center of each image (corresponding to a single band) was found using an iterative centroiding algorithm provided in the XARA pipeline. The images are then centered to the nearest pixel, before being background subtracted and cropped down to a square 65 pixel subframe. A "super-Gaussian" window function with half-width half-maximum of 20 pixels is applied to each image to smooth the Fourier domain phases. This effectively reduces noise contributions from pixels far from the PSF, leading to more robust kernel phases.

Using the XARA pipeline, phases are then extracted from each image with a discrete Fourier transform at frequencies corresponding to the selected UV-plane sampling points. Next a subpixel centroid correction is derived by conducting a grid search to find the pixel shift that minimizes the scatter in the phases. This is equivalent to fitting a plane to the phases within the optical transfer function of the Fourier transform since a shift in the image space corresponds to multiplying by a phase ramp in Fourier space. The phases are then corrected for this offset before being left multiplied by the kernel-phase matrix to derive the kernel phases. A total of 107 kernel phases are computed from each image corresponding to a single spectral bin from each data cube.

The quality of the data varies during the course of the observations, with some data cubes showing more noise than is typical. To boost the sensitivity of the analysis, these cubes can be removed. In this context, a point-source calibrator is expected to have a kernel-phase value of zero, and so any measured signal is likely due to instrumental systematics. An estimate of the level of noise can be derived by computing the standard deviation of the kernel phases across kernel-phase index. The average value of these standard deviations is then taken across each of the 17 spectral bins, producing a single standard deviation value for each data cube. For the MWC 758 observations, data cubes with an average standard deviation higher than 0.5 rad were removed. This excluded two frames from the HD 245009 data and 28 frames from the MWC 758 data including a section at the start of observations. The data quality for the MWC 480 observations was poorer, so frames with an average standard deviation across kernel phase of 1.0 rad were removed, excluding the first 220 frames and an additional 46 that were interspersed in the remaining data. The values of 0.5 and 1.0 rad were chosen by visually inspecting the standard deviations of the cubes over time. This was done with the goal of preserving as much of the data as possible, while removing clearly outlying frames. For the calibrator HD 56386, only kernel phases from the undithered frames were used, consisting of 15 of the 55 data cubes, in order to limit the risk of introducing additional systematics associated with having the image fall on a different location on the detector. All calibrator frames had a standard deviation (across kernel-phase index) of lower than 1.0 rad. Introducing these data quality checks reduces the standard deviation of the kernel phases (across kernel-phase index) from 0.57 to 0.45 rad for MWC 758 and from 1.72 to 0.65 rad for MWC 480.

---
[12] https://github.com/fmartinache/xara

In order to maximize sensitivity to candidates at higher separations and to improve the error analysis, a statistically independent set of kernel phases is constructed following the procedure set out in Ireland (2013): First, the covariance of a kernel-phase data set, $C_k$, is calculated empirically. It is also possible to compute an analytical covariance (e.g., Kammerer et al. 2019); however, here we have opted not to do this due to the high SNR of the data and good number of available frames. Next, a new kernel-phase operator $K_s$ is constructed by first diagonalizing $C_k$ using the finite dimensional spectrum theorem,

$$C_k = S^T \cdot D \cdot S, \tag{7}$$

then left multiplying the original kernel-phase matrix, $K_0$, by the unitary matrix $S$ such that

$$K_s = S \cdot K_o = S \cdot K \cdot R. \tag{8}$$

This new operator is then applied to each of the data sets in order to produce statistically independent kernel phases. The final step, prior to using the kernel phases for model fitting, is to remove any residual systematic signals that may be present in the data. While kernel phases are "self-calibrating" to first order, residual systematics are typically present in the data and often are the limiting factor for performance (e.g., Martinache et al. 2020). For this study, we calibrate the data in two ways, using a reference differential imaging (RDI) approach and a spectral differential imaging (SDI) approach.

We tailor these projections differently for the two different calibration schemes considered in this paper. These are the RDI and SDI approaches. For RDI, statistically independent kernel phases are constructed by taking the average covariance of both the science target and the PSF calibrator star for each respective data set. The same projection must be used for both the science target and the calibrator, as using a different projection may introduce systematic differences between the two sets of kernel phases. For SDI, the average covariance across all 17 spectral channels of the science target is taken.

As a result, two distinct sets of kernel phases are produced for this study, with one used with the RDI calibration and the other used with the SDI calibration. The RDI calibration is used to search for a continuum signal present in the data, for example, from a companion or disk asymmetry. Such a signal would normally be self-subtracted by the SDI calibration. The SDI approach is only able to distinguish relative changes between adjacent spectral bins in the data cube. This is ideal for identifying line absorption or emission.

In order to apply the calibration, starting with RDI, we use the respective PSF calibrators for each MWC 758 and MWC 480 and compute a mean kernel-phase value for each kernel-phase index for each spectral bin. This is computed by taking the average of all frames for the calibrator. This mean "trend" signal is then subtracted from the science target data. For the SDI calibration, we compute a "mean trend" using the average of the spectral bins adjacent to the Br-$\gamma$ bin, as detailed in Chaushev et al. (2023). However, instead of taking the mean across all frames, this is computed per pointing of 5°–6° of parallactic angle evolution. Doing so allows for better removal of short-term systematics in the data, which may vary through the course of the night, and also limits the self-subtraction of any potential science signals in the data. This is because, while any real signal in the data will change as a function of parallactic angle, it is not expected to vary much over 5°. We can then compare the calibration done "per pointing" to a "time-average" calibration where all data are binned together. For MWC 480, this consists of three consecutive pointings covering a $\Delta$PA of 17°.46. The pointings consist of 168, 142, and 126 frames, respectively. For MWC 758, the data comprised two binned pointings with a PA of 88° (54 frames) and 96° (45 frames), respectively.

It is worth pointing out that a technique such as angular differential kernel phase (ADK; Laugier et al. 2020) may be useful for this data. Laugier et al. (2020) show that it is possible to construct observables that take into account sky rotation directly while maintaining the same statistical properties as the original kernel phases. This results in smaller and more Gaussian calibration residuals, which is extremely beneficial. However, the authors mention that this may reduce sensitivity at small separations for data sets with limited field rotation. We have, therefore, opted not to apply it for this data, where our parallactic angle evolution is limited; however, ADK is an extremely promising way to improve the results of future observations using SCExAO/CHARIS and kernel phase.

## 4. Analysis and Results

Using the RDI and SDI calibrated kernel phases, we conduct a search for companions in $K$-band continuum and Br-$\gamma$ emission for both the MWC 758 and MWC 480 data sets.

### 4.1. Br-$\gamma$ Emission Search

For the analysis, we found that the "per-pointing" calibration resulted in marginally better sensitivity ($\sim$0.1–0.2 mag) than using the "time-averaged" calibration. This is suggestive of short-term systematic trends in the data, which may be the result of variable seeing or changing AO performance over the course of the observations. However, due to the limited size of the data sets, it is difficult to draw conclusions about which strategy may be the best for future observation.

Therefore, using the "per-pointing" SDI calibrated kernel phases, we conduct a search for evidence of Br-γ emission by fitting a companion model to the data. This is done in a three-step process. First, a grid of simulated binary companions is constructed for different values of each of the three fitted parameters: contrast, separation, and position angle (PA). The grid consists of 172,800 different test binaries spanning 20–80 mas in 2.5 mas increments, 360° of PA in 5° increments, and a contrast ratio ($\Delta m$) of 8 mag in increments of 0.08 mag. We find the best fit by identifying the model binary that minimizes the $\chi^2$ residual between itself and the data for each individual wavelength bin. This is done to keep the grid size manageable.

For MWC 758, this yields a solution of 20 mas and mean contrast of 5.3–6.5 mag with a range of PAs for each of the three pointings. The differences in PA between the best grid fit of the three pointings indicate an inconsistency between the solutions. For MWC 458, the grid minimization finds a similar solution of 20 mas and mean contrast of approximately 5.5 mag with a range of PAs. Examining the solutions for each pointing, there appears to be little correspondence between the observed and fit kernel phases, with the grid fit attempting to minimize the amplitude of the simulated binary kernel phases. This is consistent with the fact that the solutions at 20 mas are near to (or even below) the noise limit of the data, corresponding to significance levels ranging from $0.5\sigma$ to $1.8\sigma$. Using the grid fit as a starting point, an optimization is run to further refine the solution in the hope of improving the fit. This is done using the Nelder–Mead optimization function from the SciPy library, with the goal of reducing the $\chi^2$ residual between the data and the companion model. The optimizer finds a similar solution to the grid fit, with the parameters changing less than 10%. This improves on the fit slightly but still produces a $<2\sigma$ fit. In the absence of a good solution, the optimization algorithm may be getting stuck in a local minima or saddle point. To test this, we randomly changed the starting point of the minimization 100 times, and in each case a similarly poor (but different) solution was found.

As a final cross check, using both the grid and optimizer solutions as an initial starting point, a Markov Chain Monte Carlo (MCMC) fit is run using the emcee package to further explore the parameter space. After 10,000 steps, the MCMC fit failed to converge on a solution for both MWC 758 and MWC 480. Therefore, within the noise limits of the observations, we can reject the possibility of a Br-γ companion in both the MWC 758 and MWC 480 systems.

Finally, to estimate the noise level of the data and set upper bounds on the Br-γ line emission in the system, we use the $\chi^2$ interval method. Using the same process as in Chaushev et al. (2023), the achievable contrast is computed by comparing the $\chi^2$ interval between the data and a null model (no signal), and the data and model binaries of different contrasts, PAs, and separation. Figure 3 shows the estimated contrast for the two data sets. Two contrast curves are included representing the SDI and RDI calibrations. In both cases, the SDI methods results in the higher contrast for Br-γ, with the RDI calibration performing between 1 and 2 mag worse (depending on the angular separation). The difference in these two calibrations may be due to differences between the science target and calibrator or the presence of a real KP signal or from a close-in disk asymmetry.

For Figure 3, we compute an estimated limit of the mass times accretion rate by taking the Br-γ contrast limit and converting it into a flux limit using the stellar Br-γ flux (Eisner et al. 2009). We then assume an empirical relationship between the line luminosity and the accretion luminosity taken from Rigliaco et al. (2012). This sets a provisional upper limit on ongoing accretion in both MWC 480 and MWC 758 in the range of $10^{-5}$ to $10^{-6} M_{jup}^2 \cdot yr^{-1}$. These values are high in comparison to the PDS 70b and c, where accretion rates of $\sim 10^{-8} M_{jup} \cdot yr^{-1}$ were determined (Haffert et al. 2019). However, the accretion rates represent the end stages of planet formation and are necessarily too low to form the 4–10 $M_{jup}$ planets PDS 70b and c in the typical $\sim$3–5 Myr disk lifetime (Ribas et al. 2015; Ansdell et al. 2017). A much higher accretion rate is needed to form such massive planets. For example, assuming that PDS 70b and c are $\sim 5 M_{jup}$ and they form over 5 Myr, we would need an average accretion rate of $10^{-6} M_{jup} \cdot yr^{-1}$ to produce them. This is much more in line with our current sensitivities, and therefore, our observations can place some limits on massive ($\sim 10 M_{jup}$) rapidly accreting planets in both systems, especially since the accretion rate is expected to be higher earlier in the formation history of the planet (Drazkowska et al. 2023). In the case of the very earliest stage of formation, e.g., a planetary core undergoing runaway accretion, we would be sensitive to masses as low as $\sim 1 M_{jup}$ (Ginzburg & Chiang 2019).

### 4.2. Continuum Emission Search

While the Br-γ search resulted in a nondetection, only a single spectral band is used. By leveraging the full-wavelength information available to look for continuum emission, it is possible to increase the sensitivity of the companion search. For this, the RDI-calibrated kernel phases are used since SDI will self-subtract any steady continuum signal.

The fitting process from Section 4.1 (Br-γ) is repeated on both data sets. The grid search is conducted jointly using all wavelengths and yields a solution of 50 mas and a contrast of 7.3 with a PA of 200° for MWC 758. For MWC 480, a solution of 35 mas with a contrast of 7.2 and a PA of 15° are found. Both these contrast values are near the noise limit of the data at $0.8\sigma$ and $1.3\sigma$, respectively. The grid-fit solution is then used as a starting point for a minimization run using the Nelder–Mead optimization function available from the SciPy library. The optimizer returns a similar but refined solution. An MCMC fit is then run using 64 walkers for 10,000 steps. The walkers are initialized using the optimizer solutions with a scatter of 10% in each of the three parameters. The MCMC fails to converge to a single, well-defined minima after 10,000 steps. This is consistent with the low statistical significance of the optimizer solutions, and therefore, similarly to the Br-γ case, we can exclude any companions in the data set.

Figure 4 shows the contrast curves for the data that are created using the $\chi^2$ confidence interval method (as discussed in Section 4.1). The K-band continuum data (contours of Figure 4) have been calibrated using RDI. All solutions found by the fitting process fall outside the sensitivity range of the data. By simultaneously fitting all wavelengths, we increase the signal-to-noise by a factor of $\sim 4$, as compared to the Br-γ RDI calibration alone, gaining close to 1.5 mag in contrast. Using the COND models (Baraffe et al. 2003), we translate the contrast limit to a planet mass assuming ages of 6 and 8 Myr, respectively, for MWC 480 and MWC 758. Due to the relatively bright nature of both targets, coupled with the moderate achievable contrast, we are only able to rule out high-mass brown dwarfs ($M > 20\,M_{\rm jup}$). This precludes us from placing any new constraints on the currently existing planet candidates in the system.

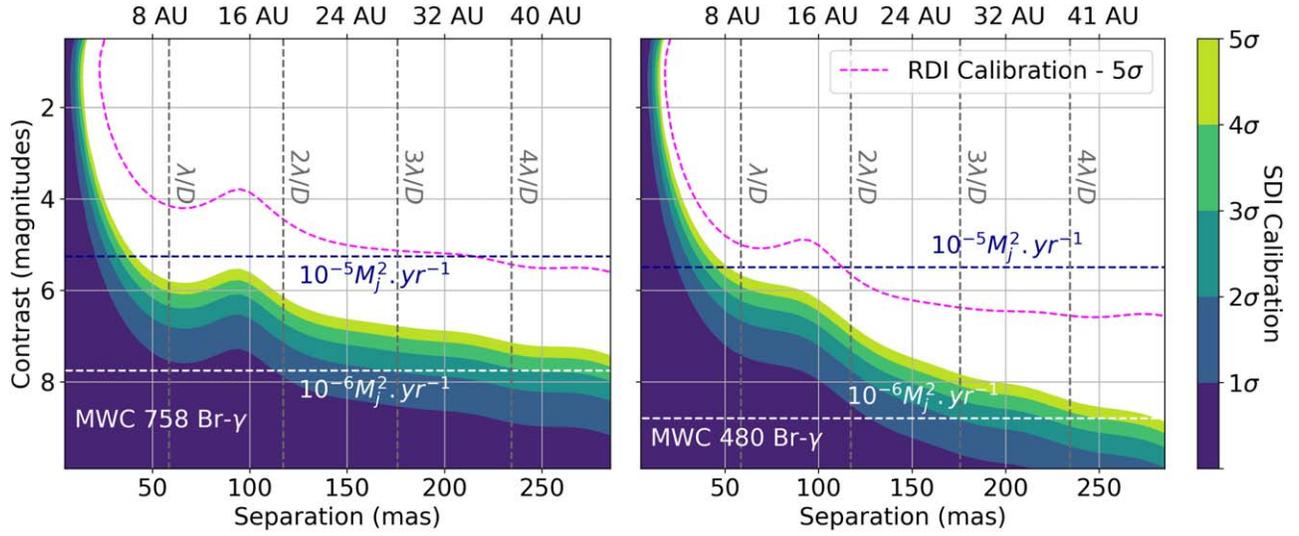

**Figure 3.** Estimated achieved contrast (contours) for band 7/2163 nm of the high-resolution *K*-band data for MWC 758 (left) and MWC 480 (right). These contours were calculated using the "per-pointing" SDI calibration and estimated by using the $\chi^2$ interval method as discussed in Chaushev et al. (2023). The equivalent sensitivity of the RDI calibration (pink dashed line) is also shown. SDI improves on the RDI sensitivity by up to ∼2 mag depending on the separation. An example typical inner working angle for a CHARIS Lyot corongraph would range anywhere from 1 to 3 $\lambda/D$. An accretion rate of $10^{-6}\,M_{\rm jup}\cdot{\rm yr}^{-1}$ would form a 5 $M_{\rm jup}$ planet on the typical disk-lifetime scale of ∼5 Myr. Therefore, assuming a planet mass of 10 $M_{\rm jup}$, we would be sensitive to accretion rates as low as $10^{-7}\,M_{\rm jup}\cdot{\rm yr}^{-1}$ at larger separations and $10^{-6}\,M_{\rm jup}\cdot{\rm yr}^{-1}$ down to around 50 mas. This sensitivity allows us to place limits on rapidly accreting massive planets around MWC 480 and MWC 758. For 1 $M_{\rm jup}$ planets, we would only be able to detect the planet during the runaway accretion phase or shortly thereafter, when accretion rates are exceptionally high (Ginzburg & Chiang 2019).

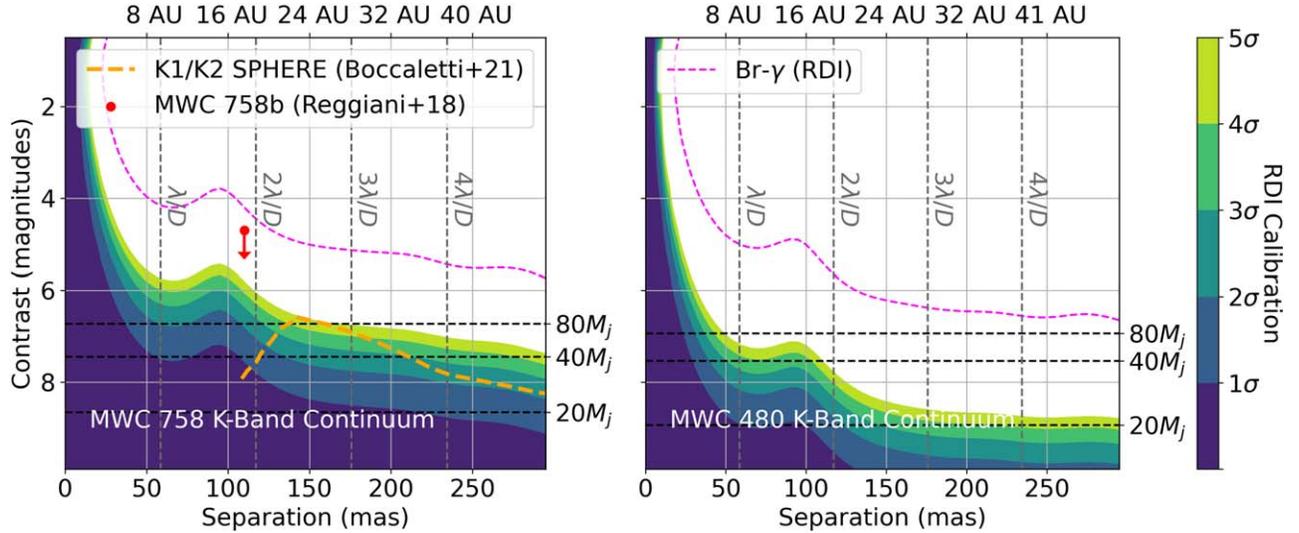

**Figure 4.** Contrast curves for the RDI-calibrated kernel-phase data sets for MWC 758 and MWC 480. The contrast curves are calculated using a $\chi^2$ interval method, which measures the difference in $\chi^2$ between the data and a range of simulated test binaries. The figures also show the contrast curve for spectral bin 7 (pink dashed line), corresponding to where we might expect to find line emission in the case of Br-$\gamma$ emission. While the SDI calibration is more sensitive than RDI for finding relative changes in the flux ratios, the RDI contrast benefits from being able to simultaneously fit all 17 spectral bins of the data. For both MWC 758 and MWC 480, lines are included to show the expected contrast of a companion for masses of 20, 40, and 80 $M_j$. These masses were computed using the COND models (Baraffe et al. 2003). We use ages of 8 and 6 Myr for MWC 758 and MWC 480, respectively (Vioque et al. 2018). Assuming that any companion was formed shortly after the star, the continuum observations are only sensitive to brown dwarf mass objects.

## 5. Conclusion

We have conducted a high-angular resolution search for protoplanets around MWC 480 and MWC 758 interior to previous studies. The search was conducted using the KPI technique in order to search for Br-$\gamma$ line emission from a rapidly accreting protoplanet. We can rule out accretion signals in Br-$\gamma$ corresponding to accretion rates of $10^{-5}$–$10^{-6}\,M_{\rm jup}^2\cdot{\rm yr}^{-1}$. Additionally, we conducted a search for signs of *K*-band continuum emission from any young massive candidates. Due to the fact that both MWC 758 and MWC 480 are bright in *K* band, using the assumed ages of 6 and 8 Myr, we can rule out massive brown dwarf companions in the system. No evidence was found either for Br-$\gamma$ emission or for a candidate brown dwarf interior to previous searches. Nevertheless, these data demonstrate the capabilities of ground-based KPI with an IFS. Future work will focus on improving the calibration strategy of this technique, the efficacy of which currently limits the ability to reach higher contrast ratios. Further development of SDI-KPI is a promising avenue for achieving this, and coupled with other approaches for better modeling the telescope systematics (e.g., Pope et al. 2021), this may deliver a significant boost in contrast. Additionally, this technique may be applicable to JWST IFUs, and in particular the NIRSpec IFU mode, enabling moderate-resolution spectroscopy of close-in protoplanets from space. The longer-wavelength coverage and higher SNR provided by JWST would allow for detailed characterization as well as sensitivity to less massive protoplanets.

### Acknowledgments

We would like to thank the anonymous reviewer for their careful reading of our manuscript and their insightful comments and suggestions, which have substantially improved this paper. This research was funded by the Heising-Simons Foundation through grant 2020-1825. Based on data collected at Subaru Telescope, which is operated by the National Astronomical Observatory of Japan. The development of SCExAO is supported by the Japan Society for the Promotion of Science (Grant-in-Aid for Research #23340051, #26220704, #23103002, #19H00703, #19H00695 and #21H04998), the Subaru Telescope, the National Astronomical Observatory of Japan, the Astrobiology Center of the National Institutes of Natural Sciences, Japan, the Mt. Cuba Foundation and the Heising-Simons Foundation. The authors wish to recognize and acknowledge the very significant cultural role and reverence that the summit of Maunakea has always had within the indigenous Hawaiian community, and are most fortunate to have the opportunity to conduct observations from this mountain. This research made use of POPPY, an open-source optical propagation Python package originally developed for the James Webb Space Telescope project (Perrin et al. 2012). This research has also made use of the following scientific packages: *NumPy* (Harris et al. 2020), *Matplotlib* (Hunter 2007) and *SciPy* (Virtanen et al. 2020).

### ORCID iDs

Alexander Chaushev https://orcid.org/0000-0003-0061-5446
Steph Sallum https://orcid.org/0000-0001-6871-6775
Julien Lozi https://orcid.org/0000-0002-3047-1845
Tyler Groff https://orcid.org/0000-0001-5978-3247
Olivier Guyon https://orcid.org/0000-0002-1097-9908
Andy Skemer https://orcid.org/0000-0001-6098-3924

### References

Ansdell, M., Williams, J. P., Manara, C. F., et al. 2017, AJ, 153, 240
Baraffe, I., Chabrier, G., Barman, T. S., Allard, F., & Hauschildt, P. H. 2003, A&A, 402, 701
Betti, S. K., Follette, K. B., Ward-Duong, K., et al. 2022, ApJL, 935, L18
Boccaletti, A., Pantin, E., Ménard, F., et al. 2021, A&A, 652, L8
Bonneau, D., Clausse, J. M., Delfosse, X., et al. 2006, A&A, 456, 789
Bonneau, D., Delfosse, X., Mourard, D., et al. 2011, A&A, 535, A53
Brandt, T. D., Rizzo, M., Groff, T., et al. 2017, JATIS, 3, 048002
Chaushev, A., Sallum, S., Lozi, J., et al. 2023, JATIS, 9, 28004
Chelli, A., Duvert, G., Bourgès, L., et al. 2016, A&A, 589, A112
Drazkowska, J., Bitsch, B., Lambrechts, M., et al. 2023, in ASP Conf. Ser. 534, Protostars and Planets VII, ed. S. Inutsuka et al. (San Francisco, CA: ASP), 717
Eisner, J. A. 2015, ApJL, 803, L4
Eisner, J. A., Graham, J. R., Akeson, R. L., & Najita, J. 2009, ApJ, 692, 309
Factor, S. M., & Kraus, A. L. 2023, AJ, 165, 130
Gaia Collaboration, Brown, A. G. A., Vallenari, A., et al. 2018, A&A, 616, A1
Ginzburg, S., & Chiang, E. 2019, MNRAS, 490, 4334
Gravity+ Collaboration, Abuter, R., Alarcon, P., et al. 2022, Msngr, 189, 17
Groff, T., Chilcote, J., Kasdin, N., et al. 2016, Proc. SPIE, 9908, 990800
Groff, T. D., Kasdin, N. J., Limbach, M. A., et al. 2015, Proc. SPIE, 9605, 457
Haffert, S. Y., Bohn, A. J., de Boer, J., et al. 2019, NatAs, 3, 749
Harris, C. R., Millman, K. J., van der Walt, S. J., et al. 2020, Natur, 585, 357
Hunter, J. D. 2007, CSE, 9, 90
Ireland, M. J. 2013, MNRAS, 433, 1718
Kammerer, J., Cooper, R. A., Vandal, T., et al. 2023, PASP, 135, 014502
Kammerer, J., Ireland, M. J., Martinache, F., & Girard, J. H. 2019, MNRAS, 486, 639
Kammerer, J., Kasper, M., Ireland, M. J., et al. 2021, A&A, 646, A36
Keppler, M., Benisty, M., Müller, A., et al. 2018, A&A, 617, A44
Laugier, R., Martinache, F., Cvetojevic, N., et al. 2020, A&A, 636, A21
Liu, Y., Dipierro, G., Ragusa, E., et al. 2019, A&A, 622, A75
Luhman, K. L. 2023, AJ, 165, 37
Martinache, F. 2010, ApJ, 724, 464
Martinache, F., Ceau, A., Laugier, R., et al. 2020, A&A, 636, A72
Müller, A., Keppler, M., Henning, T., et al. 2018, A&A, 617, L2
Nielsen, E. L., Rosa, R. J. D., Macintosh, B., et al. 2019, AJ, 158, 13
Perera, S., Pott, J.-U., Woillez, J., et al. 2022, MNRAS, 511, 5709
Perrin, M. D., Soummer, R., Elliott, E. M., Lallo, M. D., & Sivaramakrishnan, A. 2012, Proc. SPIE, 8442, 84423D
Pope, B., Martinache, F., & Tuthill, P. 2013, ApJ, 767, 110
Pope, B. J. S., Pueyo, L., Xin, Y., & Tuthill, P. G. 2021, ApJ, 907, 40
Ribas, Á., Bouy, H., & Merín, B. 2015, A&A, 576, A52
Reggiani, M., Christiaens, V., Absil, O., et al. 2018, A&A, 611, A74
Ren, B., Dong, R., van Holstein, R. G., et al. 2020, ApJL, 898, L38
Rigliaco, E., Natta, A., Testi, L., et al. 2012, A&A, 548, A56
Ruane, G., Ngo, H., Mawet, D., et al. 2019, AJ, 157, 118
Sallum, S., Eisner, J., Skemer, A., & Murray-Clay, R. 2023, ApJ, 953, 55
Sallum, S., & Skemer, A. 2019, JATIS, 5, 018001
Teague, R., Bae, J., Aikawa, Y., et al. 2021, ApJS, 257, 18
Tuthill, P. G., Monnier, J. D., Danchi, W. C., Wishnow, E. H., & Haniff, C. A. 2000, PASP, 112, 555
Vioque, M., Oudmaijer, R. D., Baines, D., Mendigutía, I., & Pérez-Martínez, R. 2018, A&A, 620, A128
Virtanen, P., Gommers, R., Oliphant, T. E., et al. 2020, NatMe, 17, 261
Wagner, K., Stone, J., Skemer, A., et al. 2023, NatAs, 7, 1208
Wagner, K., Stone, J. M., Spalding, E., et al. 2019, ApJ, 882, 20
Widmann, F., Pott, J.-U., & Velasco, S. 2018, MNRAS, 475, 1224
Zhu, Z. 2015, ApJ, 799, 16